\documentclass[twocolumn]{jpsj3}
\usepackage{amsmath,amssymb,mathrsfs}
\usepackage{graphicx}
\usepackage{overpic}
\usepackage{wrapfig}
\usepackage{bm}
\usepackage{times}
\usepackage{txfonts}

\usepackage[usenames]{color}

\def\gsim{\; $\raise0.3ex\hbox{$>$}\llap{\lower0.8ex\hbox{$\sim$}}$\;}
\def\lsim{\; $\raise0.3ex\hbox{$<$}\llap{\lower0.8ex\hbox{$\sim$}}$\;}

\title{Spin-Wave Analysis for Kagome-Triangular Spin System and Coupled Spin Tubes:
Low-Energy Excitation for the Cuboc Order} 
\author{Masahiro Ochiai,  Kouichi Seki, and Kouichi Okunishi$^1$}
\inst{Graduate School of Science and Technology, Niigata University, Niigata 950-2181, Japan \\
$^1$Department of Physics, Niigata University, Niigata 950-2181, Japan}
\abst{
The coupled spin tube system, which is equivalent to a stacked Kagome-triangular spin system, exhibits the cuboc order -- a non-coplanar spin order with a twelve-sublattice structure accompanying spontaneous breaking of the translational symmetry -- in the Kagome-triangular plane.
On the basis of the spin-wave theory, we analyze spin-wave excitations of the planar Kagome-triangular spin system, where the geometric phase characteristic to the cuboc spin structure emerges.
We further investigate spin-wave excitations and dynamical spin structure factors for  the coupled spin tubes, assuming the staggered cuboc order.
}
\kword{coupled spin tubes, Kagome-triangular lattice, Cuboc order, spin wave}
\begin{document}
\maketitle

\section{Introduction}

Frustration effects in spin systems often induce nontrivial spin orders,  such as a 120$^\circ$ structure with spin chirality or a non-coplanar spin order, accompanying spontaneous breaking of the  translation and spin rotational symmetries.
Among various interesting spin orders, the cuboc order formed in a quasi-two-dimensional (quasi-2D) plane with frustrating interactions has recently attracted much interest.\cite{Domenge,Domenge2,Messio} 
The cuboc order is defined as a non-coplanar spin order with a twelve-sublattice structure, which can be specified by a triple-wavevector structure in the momentum space.
Recent experiments on Kagome-lattice-based spin systems such as NaBa$_2$Mn$_3$F$_{11}$\cite{Ishikawa} and  Cu$_3$Zn(OH)$_6$Cl$_2$\cite{Kapella} actually suggest the possibility of the cuboc order.
Moreover, the phase transition associated with the cuboc order provides  fascinating physics from the theoretical viewpoint.\cite{Seki}

\begin{figure}[tb] 
\centering\includegraphics[width=7cm]{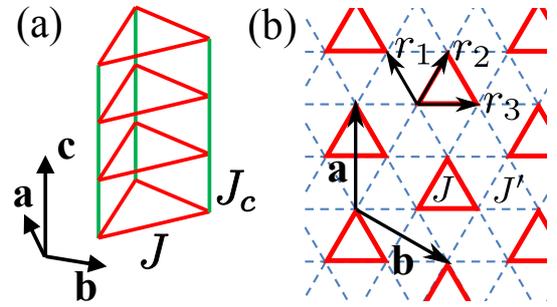}
\caption{(Color online)
(a) Lattice structure of a spin tube. $J$ denotes the exchange coupling in the unit triangle and $J_c$ is the coupling in the tube-leg direction along the $c$-axis.
(b) Kagome-triangular lattice structure of coupled spin tubes in the $ab$-plane.
Spins on triangles of the tubes are coupled with the inter-tube coupling $J'$ (broken lines).
The inter-tube coupling of $J'$ has a Kagome lattice structure, whereas the intra-tube couplings of $J$ (triangles of solid lines) correspond to  next-nearest couplings of the Kagome lattice. 
The lattice translation vectors in the $ab$-plane are defined as ${\bf a}$ and ${\bf b}$, while  $\bm r_1$, $\bm r_2$, and $\bm r_3$ denote the vectors indicating the nearest-neighbor sites.
}\label{fig1}
\end{figure}

Another interesting compound for the cuboc order is the coupled spin tube system CsCrF$_4$.\cite{CsCrF4_Manaka_2009,CsCrF4_Manaka_2011}
AC-susceptibility and neutron diffraction experiments\cite{ManakaAC,Hagihala} suggest signatures of  nontrivial spin order below $T<4$ K.
Nevertheless, the order realized in CsCrF$_4$ has not been experimentally specified yet.
Then, an essential point for CsCrF$_4$ is that a small but non-negligible inter-tube coupling forms the Kagome-triangular lattice\cite{KTlattice} in a cutting plane of the coupled spin tubes (see Fig. \ref{fig1}).
Since the exchange coupling in the tube-leg direction is not frustrating, the Kagome-triangular lattice structure plays a significant role in forming the cuboc order. 
Monte Carlo simulations for the classical Heisenberg model of spin tubes coupled with the weak ferromagnetic inter-tube coupling have demonstrated that the cuboc order actually emerges at a finite temperature.\cite{Seki}
However, the physical properties (involving quantum effect) of the cuboc order for the coupled spin tubes have not been quantitatively investigated yet.

In this paper, using the spin-wave theory, we study low-energy excitations of the coupled spin tubes, or, equivalently, of a stacked Kagome-triangular system.
We first analyze dispersion relations of the spin-wave Hamiltonian for the 2D Kagome-triangular plane.
Then,  a key point  is that the spin quantization axes are nontrivially tilted in the cuboc spin configuration, for which the hopping matrix elements may acquire  geometric phases.
We next investigate spin-wave excitations of the coupled spin tubes with 3D couplings, assuming the  staggered cuboc order in the $c$-axis (tube-leg) direction.
In particular, we calculate the dynamical spin structure factors for the coupled spin tubes in addition to the spin-wave dispersion relations.
On the basis of these spin-wave results, we clarify features of the spin-wave excitations for the coupled spin tubes with the cuboc order and discuss their relevance to CsCrF$_4$.

This paper is organized as follows.
In the next section, we summarize basic properties of the coupled spin tubes and the cuboc order.
In \S 3, we analyze the spin-wave Hamiltonian for the 2D Kagome-triangular plane, which is represented as a $24 \times 24$ matrix reflecting the twelve sublattice structure of the cuboc order.
In \S 4, we perform the spin-wave analysis of the coupled spin tubes with the full 3D interactions, assuming the staggered cuboc order in the tube-leg direction.
In \S 5, we summarize our results and then discuss their relevance to  recent experiments.

\section{Coupled Spin Tubes and Cuboc Order}

Let us introduce the coupled spin tube model. 
As shown in Fig. \ref{fig1}(a), we assume that the unit triangle of a spin tube is located in the $ab$-plane and the tube-leg direction is along the $c$-axis.
Then, the Hamiltonian of the coupled spin tubes is written as
\begin{equation}
{\cal H} = J\sum_{\langle i,j\rangle_{\rm intra}} {\bm S}_i\cdot {\bm S}_j + J' \sum_{\langle i,j\rangle_{\rm inter}} {\bm S}_i\cdot {\bm  S}_j  +  J_c\sum_{\langle i,j\rangle_{\rm leg}} {\bm S}_i\cdot {\bm S}_j ,
\label{chamiltonian}
\end{equation}
where $\bm S$ denotes a vector spin for the classical case or  a spin matrix with  magnitude $S$ for the quantum spin.
$J$ and $J_c$ respectively represent the exchange couplings in the unit triangle and in the tube-leg direction.
Note that for CsCrF$_4$ the intra-tube and tube-leg couplings are antiferromagnetic and $J_c \sim 2 J \gg |J'|$ is expected\cite{Koo}.
Although interesting ground-state properties of the single $S=1/2$ quantum spin tube have been clarified so far\cite{Kawano,Cabra,Arikawa,Spintube1,Spintube2}, here, we emphasize the importance of the small but finite inter-tube coupling $J'$ in the context of the cuboc order.
As shown in Fig. \ref{fig1}(b), the coupled spin tubes  basically have a triangular lattice structure in the $ab$-plane.
However, an interesting aspect is that the lattice structure  of the inter-tube coupling is topologically equivalent to the Kagome lattice, and the intra-tube coupling corresponds to a next-nearest-neighbor interaction of the Kagome lattice. 
We therefore call the lattice structure of the coupled spin tubes in the $ab$-plane the Kagome-triangular lattice\cite{KTlattice}. 
This Kagome-triangular lattice structure is essential for the cuboc order, while the tube-leg coupling $J_c$ basically causes no frustration.

\begin{figure}[tb]
\centering\includegraphics[width=5cm]{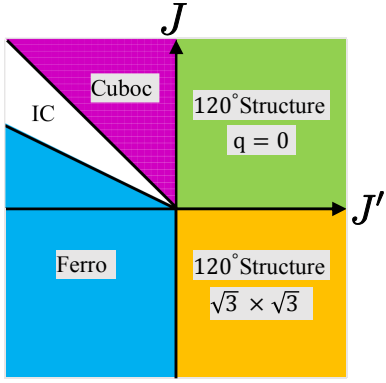}
\caption{(Color online)
Ground-state phase diagram of the classical Heisenberg model on the coupled tube lattice. The ground state basically depends only on the couplings $J$ and $J'$ in the $ab$-plane since $J_c$ causes no frustration effect.
}\label{fig2}
\end{figure}

The ground-state phase diagram of the coupled spin tubes based on the classical Heisenberg model is shown in Fig. \ref{fig2}.
Note that CsCrF$_4$ is located around $J'\sim 0$ with the antiferromagnetic intra-tube coupling $J>0$, but the determination of $J'$ is an experimentally subtle problem. 
In this paper, we basically assume the ferromagnetic inter-tube coupling, $J'<0$, and particularly focus on the weak inter-tube coupling region, $-1<J'/J<0$, where the cuboc order appears.
For the classical coupled spin tubes, the cuboc order has actually been confirmed to be realized at a finite temperature by Monte Carlo simulations.\cite{Seki}
As $|J'/J|$ increases, the incommensurate order emerges in $-2<J'/J<-1$, and finally the ferromagnetic order becomes stable for $J'/J<-2$.

\begin{figure}[tb]
\centering\includegraphics[width=8cm]{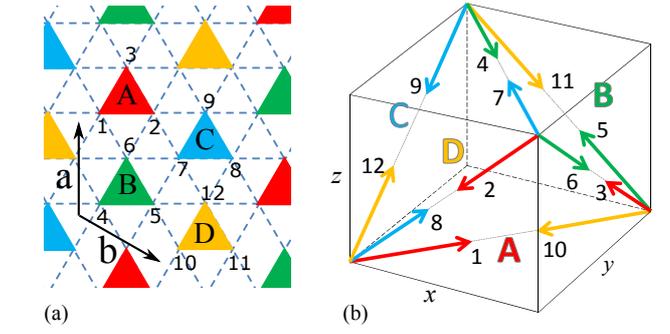}
\caption{(Color online)
Spin structure of the cuboc order.
(a) Twelve-sublattice structure of the cuboc order, where the translational symmetry is broken by $2\times 2$ in the unit of the tube triangle.
Then, the 120$^\circ$ structures defined for the triangles labeled by A, B, C, and D form a tetrahedron in the spin space, as shown in (b).
The numbers assigned for the 12 sublattices in (a) correspond to those of the spins in (b).  
$x,y$, and $z$ assigned to the edges of the box frame in (b) denote the directions of the unit vectors in the spin space.
}\label{fig3}
\end{figure}

The cuboc order is defined as a non-coplanar spin order with a twelve-sublattice structure, as in Fig. \ref{fig3}.
The strong antiferromagnetic intra-tube coupling basically imposes a rigid planar 120$^\circ$ structure on the unit triangles of the tubes.
However,  the ferromagnetic inter-tube coupling  also causes frustration.
In order to reduce the energy due to the inter-tube coupling, the rigid 120$^\circ$ planes tilt with each other, accompanying the translational-symmetry breaking of $2\times 2$ in the triangle unit.
Then, the four tilting 120$^\circ$ planes labeled by {A}, {B}, {C}, and {D} form a tetrahedron in the spin space, as depicted in Fig. \ref{fig3}(b).
Here, the number indices for the 12 spins in Fig. \ref{fig3}(a)  respectively correspond to those for spins on the tetrahedron in Fig. \ref{fig3}(b).
Another important feature of the cuboc order is that spin chiralities can be defined for the tetrahedron;
the spins on each surface triangle of the tetrahedron have a vector spin chirality  pointing to the center or the outer sides of the tetrahedron.
For example,  the vector spin chirality in the case of Fig. \ref{fig3}(b)  has ``+''.
Meanwhile, the scalar spin chirality can also be  defined for the three spins at each vertex of the tetrahedron.

The spin configuration of the cuboc order can be explicitly represented as 
\begin{align}
&\bm S_\mu (\bm r)=\cos\left(\frac{1}{2}\bm q_b\cdot \bm r\right)\frac{\bm e_x}{\sqrt 2}-\cos\left(\frac{1}{2}\bm q_a \cdot \bm r\right)\frac{\bm e_y}{\sqrt 2},\nonumber\\
&\bm S_\nu (\bm r)=\cos\left(\frac{1}{2}(\bm q_a - \bm q_b)\cdot \bm r\right)\frac{\bm e_z}{\sqrt 2}-\cos\left(\frac{1}{2}\bm q_b\cdot \bm r\right)\frac{\bm e_x}{\sqrt 2},\nonumber\\
&\bm S_\eta(\bm r)=\cos\left(\frac{1}{2}\bm q_a\cdot \bm r\right)\frac{\bm e_y}{\sqrt 2}-\cos\left(\frac{1}{2}(\bm q_a-\bm q_b) \cdot \bm r\right)\frac{\bm e_z}{\sqrt 2},
\label{equ_cuboc}
\end{align}
where $\bm q_a$ and $\bm q_b$ are the reciprocal vectors in the $ab$-plane.
The vectors $\bm e_x$, $\bm e_y$, and $\bm e_z$ represent the unit vectors in the spin space along the edges of the box frame in Fig. \ref{fig3}(b) and  the vector spin is normalized as $|S|=1$.
In addition,  the suffixes of the spins in Eq. (\ref{equ_cuboc}) take $\mu =\{ 1, 4, 7, 10 \}$,  $\nu =\{ 2, 5, 8, 11 \}$, and $\eta = \{3, 6, 9, 12\} $, depending on which sublattice point the position vector $\bm r$ indicates in Fig. \ref{fig3}(a). 
An important feature of the cuboc order is that it has the triple-$\bm q$ structure, and then the Fourier transformation of the classical Hamiltonian (\ref{chamiltonian}) has the energy minimum of the ground state at the M point in the momentum space.
 This triple-$\bm q$ structure plays an essential role in analyzing  soft modes of spin-wave excitations.

\section{Spin-Wave Analysis for the Kagome-Triangular Plane}
\label{sec_planar}

In general,  3D couplings are necessary for realizing a finite-temperature order with  spontaneous breaking of the spin rotational symmetry.
For the coupled spin tubes, however, the competition between $J$ and $J'$ on the Kagome-triangular plane plays a central role in the formation of the cuboc order.
Thus, we first consider the Hamiltonian for the 2D Kagome-triangular plane,
\begin{align}
{\cal H}_{\rm 2D} = J\sum_{\langle i,j \rangle_{\rm intra}} \bm{S}_i \cdot \bm{S}_j + J'\sum_{\langle i,j \rangle_{\rm inter}} \bm{S}_i \cdot \bm{S}_j ,
\label{H2D}
\end{align}
for which we basically discuss ground-state properties.
Here, $\bm S$ in Eq. (\ref{H2D}) is a quantum spin with the magnitude $S$.

For the construction of the spin-wave Hamiltonian, we should carefully take account of the twelve-sublattice structure; 
the spin quantization axis for each spin embedded in the twelve-sublattice structure should be locally defined along the direction of the classical spin arrow in Fig. {\ref{fig3}(b). 
Then, we perform the Holstein-Primakoff transformation for the spin of each quantization axis, and rewrite it as the unified spin coordinate defined by $\bm e_x$, $\bm e_y$, and $\bm e_z$, corresponding to the edges of the box frame in Fig. \ref{fig3}(b).
The resulting spin-wave Hamiltonian is written as
\begin{align}
{\cal H}_{\rm 2D} = {\cal H}_{0} +  {\cal H}_S + {\cal O}(S^{1/2}),
\label{2dH}
\end{align}
where ${\cal H}_{0}\equiv -6(J-2J')S^2 N$ is the classical energy and ${\cal H}_S $ denotes the linear-spin-wave Hamiltonian.
Here,  $N$ denotes the total number of  magnetic unit cells in the system.
The linear spin-wave term can be represented as a standard matrix form, 
\begin{align}
{\cal H}_S = \frac{S}{8} \sum_{\bm q} X^\dagger_{\bm q}{\rm H}_{\bm q} X_{\bm q} ,
\label{2dHl}
\end{align}
where $X_{\bm q}^{\dagger}$($X_{\bm q}$) is a vector array containing 24 boson operators and ${\rm H}_{\bm q}$ is a $24\times 24$ matrix representing the hopping amplitudes of the bosons.

The array of boson operators is explicitly defined as 
\begin{align}
X_{\bm q}^{\dagger} \equiv ( A_{\bm q}^{\dagger} \,,\, B_{\bm q}^{\dagger} \,,\, C_{\bm q}^{\dagger} \,,\, D_{\bm q}^{\dagger} \,,\, A_{-\bm q}^t \,, B_{-\bm q}^t \,, C_{-\bm q}^t \,, D_{-\bm q}^t), 
\end{align}
where $A^\dagger_{\bm q}, B^\dagger_{\bm q}, C^\dagger_{\bm q}$, and  $D^\dagger_{\bm q}$  respectively represent a set of boson creation operators carrying momentum $\bm q$ for the unit triangle spins labeled by A, B, C, and D in Fig. \ref{fig3}.
For instance, the operators $A^\dagger_{\bm q}$$(A_{-\bm q}^t)$ contain three boson operators, 
\begin{align}
A_{\bm q}^{\dagger} = (a_{1, \bm q}^{\dagger} \,, a_{2, \bm q}^{\dagger}\,, a_{3, \bm q}^{\dagger} ), \;
 A_{-\bm q}^t = (a_{1, -\bm q} \,, a_{2, -\bm q} \,, a_{3, -\bm q}),
\end{align}
where $a^\dagger_{\nu , \bm q}~(a_{\nu , \bm q})$ is the boson creation (annihilation) operator in the momentum space with the sublattice index $\nu=\{1,2,3\}$ for the unit triangle labeled by A.\cite{subindex}
 Explicitly, we have
$a^\dagger_{\nu , \bm q} = \frac{1}{N} \sum_{\bm r}a^\dagger_{\nu}(\bm r) e^{-i\bm q\cdot \bm r}$,
where $\bm r$ is a lattice translation vector in the unit of the magnetic unit cell with the 12 sublattices, and the momentum $\bm q$ runs in the magnetic Brillouin zone.

For the above array of boson operators, we can write  the transition amplitude  matrix as  
\begin{align}
{\rm H}_{\bm q} \equiv
\begin{pmatrix}
P_{\bm q} & Q_{\bm q} \\
Q_{\bm q}^{\dagger} & \bar{P}_{\bm q}
\end{pmatrix}  ,
\end{align}
where $P_{\bm q} $ and $Q_{\bm q}$ denote $12 \times 12$ matrices.
Straightforward calculations lead to
\begin{align}
P_{\bm q}
&=
\begin{pmatrix}
K + \alpha_{\bm q} & -3\gamma_{1 , \bm q} & -3\gamma_{2 , \bm q} & -3\gamma_{3 , \bm q} \\
-3\gamma_{ 1, \bm q} & K + \alpha_{\bm q} & -3\gamma_{3 , \bm q} & -3\gamma_{2 , \bm q} \\
-3\gamma_{ 2, \bm q} & -3\gamma_{3 , \bm q} & K + \alpha_{\bm q} & -3\gamma_{1 , \bm q} \\
-3\gamma_{ 3, \bm q} & -3\gamma_{2 , \bm q} & -3\gamma_{1 , \bm q} & K + \alpha_{\bm q}
\end{pmatrix} ,
\label{Pmatrix} \\
\bar{P}_{\bm q}
&=
\begin{pmatrix}
K + \alpha_{\bm q} & -3\bar{\gamma}_{1 , \bm q} & -3\bar{\gamma}_{2 , \bm q} & -3\bar{\gamma}_{3, \bm q} \\
-3\bar{\gamma}_{1 , \bm q} & K + \alpha_{\bm q} & -3\bar{\gamma}_{3 , \bm q} & -3\bar{\gamma}_{2 , \bm q} \\
-3\bar{\gamma}_{2 , \bm q} & -3\bar{\gamma}_{3 , \bm q} & K + \alpha_{\bm q} & -3\bar{\gamma}_{1 , \bm q} \\
-3\bar{\gamma}_{3 , \bm q} & -3\bar{\gamma}_{2 , \bm q} & -3\bar{\gamma}_{1 , \bm q} & K + \alpha_{\bm q}
\end{pmatrix} ,
\label{Pbarmatrix} \\
Q_{\bm q}
&=
\begin{pmatrix}
-3\bar{\alpha}_{\bm q} & \eta_{1 , \bm q} & \eta_{2 , \bm q} & \eta_{3 , \bm q} \\
\eta_{1 , \bm q} & -3\bar{\alpha}_{\bm q} & \eta_{3 , \bm q} & \eta_{2 , \bm q} \\
\eta_{2 , \bm q} & \eta_{3 , \bm q} & -3\bar{\alpha}_{\bm q} & \eta_{1 , \bm q} \\
\eta_{3 , \bm q} & \eta_{2 , \bm q} & \eta_{1 , \bm q} & -3\bar{\alpha}_{\bm q}
\end{pmatrix} ,
\label{Qmatrix}
\end{align}
where $K$, $\alpha_{\bm q}(\bar{\alpha}_{\bm q})$, $\gamma_{i , \bm q}(\bar{\gamma}_{i , \bm q})$, and $\eta_{i  , \bm q}$ are $3\times 3$ matrices of the spin-wave hoppings in the unit of triangles.
Noticing the explicit forms of these matrices in Appendix A,  we briefly describe their meanings.
First, $K$ contains the diagonal coupling independent of $\bm q$, which corresponds to a chemical potential for the spin-wave bosons.
Secondly, $\alpha_{\bm q}$ represents the hopping of the  bosons within the unit triangle having the 120$^\circ$ structure on the tetrahedron in Fig. \ref{fig3}(b).
Thirdly, the $\gamma$ and $\bar{\gamma}$ matrices represent the hopping of the bosons between two adjacent triangles among A, B, C, and D in the tetrahedron.
Finally, $\bar{\alpha}_{\bm q}$ and $\eta$ describe the matrix elements for $a_{\nu, \bm q}^\dagger a_{\mu,\bm q}^\dagger$, $a_{\mu,\bm q }^\dagger b_{\nu, \bm q}^\dagger\, \cdots$,  or $a_{\mu,\bm q} a_{\nu,\bm q},\cdots$, and so forth, with $\mu\ne\nu$.

For the non-coplanar spin order, an interesting point is that the spin-wave hopping terms often acquire a geometric phase attributed to rotations of quantization axes.
For the cuboc order, the geometric phase is actually included as $e^{\pm i \phi}$ in $\bar{\alpha}_{\bm q}$, $\gamma_{\nu , \bm q}$ and $\eta_{\nu , \bm q}$.
Explicitly,  we have 
\begin{align}
\phi=\arctan(2\sqrt{2}),
\end{align}
which corresponds to the relative angle between two surface triangles on the tetrahedron in Fig. \ref{fig3}(b).
This geometric phase in the spin-wave Hamiltonian induces a nontrivial quantum effect on spin-wave dispersion relations.

\subsection{Spin-wave dispersion relation}

We diagonalize the spin-wave Hamiltonian (\ref{2dHl}) with the Bogoliubov transformation,
\begin{align}
X_{\bm q} = {\rm T}_{\bm q} X'_{\bm q} \, ,
\end{align}
where ${\rm T}_{\bm q}$ is a $24\times 24$ transformation matrix and  $X'_{\bm q}$ represents a set of new boson operators diagonalizing the spin-wave Hamiltonian.
Following  a general formulation in Ref. \citen{bogoliubov}, we can construct the Bogoliubov transformation matrix ${\rm T}_{\bm q}$ through the eigenvalue equation
\begin{align}
 g {\rm H}_{\bm q} {\rm T}_{\bm q }=  {\rm T}_{\bm q} g {\rm \Lambda}_{\bm q},
\end{align}
where  $\Lambda_{\bm q}$ is the eigenvalue matrix whose entry is $\lambda_{\nu, \bm q}$ and
\begin{align}
g \equiv
\begin{pmatrix}
I & 0 
\\
0 & -I
\end{pmatrix}  
\end{align}
specifies the commutators of bosons.
Here, $I$ is the  $12 \times 12$ identity matrix.
Taking account of the contribution of $\pm \bm q$ in Eq. ({\ref{2dHl}), we obtain the spin-wave spectrum as $\omega_{\nu,{\bm q}} \equiv 2 \lambda_{\nu,{\bm q}}$. 
We numerically diagonalize the asymmetric matrix $g{\rm H}_{\bm q}$ with {\tt ZGEEV} of LAPACK to obtain $\omega_{\nu, \bm q}$ for Eq. (\ref{2dHl}).
Also, the transformation matrix ${\rm T}_{\bm q}$ can be basically obtained as the eigenvectors of $g{\rm H}_{\bm q}$. 
Nevertheless, the eigenvectors computed under the normalization convention of {\tt ZGEEV} should be renormalized, so as to satisfy ${\rm T}_{\bm q}^{} g {\rm T}_{\bm q}^\dagger =g$.

\begin{figure}[bt]
\centering\includegraphics[width=4cm]{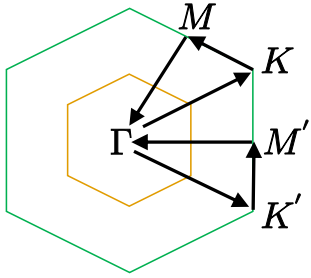}
\caption{(Color online)
Schematic diagram of the Brillouin zone.
The inner hexagon indicates the magnetic Brillouin zone for the twelve-sublattice structure of the cuboc order, while the outer hexagon represents the Brillouin zone for the Kagome-triangular plane defined by Fig. \ref{fig1}(b).
The arrows define the momentum path for spin-wave dispersion relations.
}\label{fig4}
\end{figure}

Before presenting numerical results, we comment on the Brillouin zone for the cuboc order in the Kagome-triangular plane.
In Fig. \ref{fig4}, the inner hexagon represents the reduced Brillouin zone for the cuboc order with the broken translation symmetry, whereas  the outer hexagon represents the Brillouin zone for the Kagome-triangular lattice.
In the following, we basically adopt the outer Brillouin zone of the original Kagome-triangular lattice defined by Fig. \ref{fig1}(a).
In particular, we use the path $\Gamma\to{\rm K}\to{\rm M} \to \Gamma\to {\rm K}'\to{\rm M}'\to\Gamma $ in Fig. \ref{fig4}  to show spin-wave dispersion relations, for which the $\Gamma$, M and M$'$ points are equivalent.
Here, it should be noted that the K$'$ point is not equivalent to the K point, reflecting the three fold axis symmetry of the Kagome-triangular plane.

\begin{figure}[bt]
\centering\includegraphics[width=6.1cm]{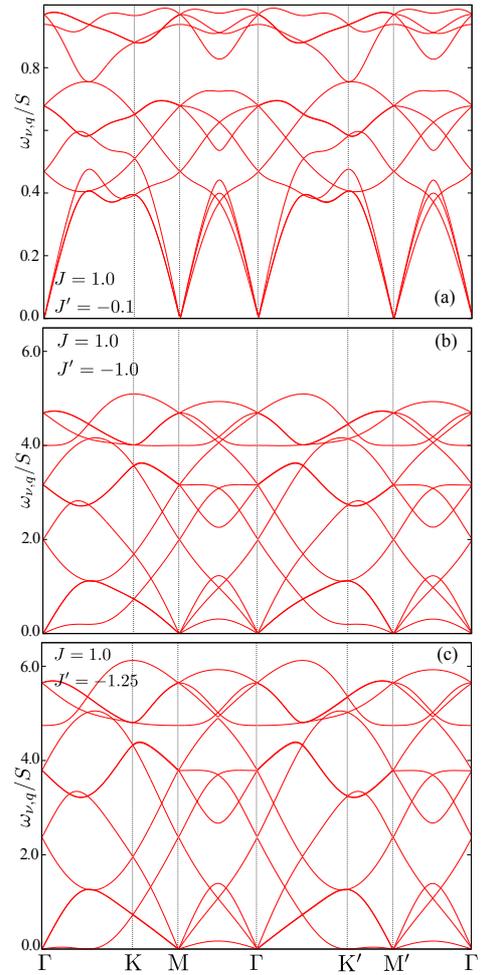}
\caption{(Color online)
Spin-wave dispersion relations for the planar Kagome-triangular system:
(a) $J=1.0$ and $J'=-0.1$, (b) $J=1.0$ and $J'=-1.0$,  (c) $J=1.0$ and $J'=-1.25$.
The horizontal axis is along the momentum path defined in Fig. \ref{fig4}.
Note that the scale of the vertical axis in (a) is different from those in (b) and (c).
}\label{fig5}
\end{figure}

In Fig. \ref{fig5}, we plot the spin-wave dispersion relations for $J'/J=-0.1$, $-1.0$, and $-1.25$.
In Fig. \ref{fig6}, we also show the lowest-energy excitation in the 2D momentum space for $J=-J'=1.0$.
In these figures, we can see that the spin-wave excitations have soft modes at the $\Gamma$, M,  and M$'$ points, reflecting the triple-$\bm q$ structure of Eq. (\ref{equ_cuboc}) for the cuboc order.
As $|J'|$ increases, the asymmetry between the K and K$'$ points becomes significant.
A higher-energy mode at the K$'$ point gradually decreases, while the modes at the K points maintain $\omega\sim 0.5$ in Figs. \ref{fig5}(b) and (c).
At $J'/J=-1.25$, we find that the lowest-energy branch at the K$'$ point touches  $\omega=0$, where its low-energy behavior turns out to be $\omega_{\bm q}\sim (\bm q-\bm q_{K'})^2$.
This behavior suggests that the cuboc order is destabilized at $J'/J=-1.25$ within  the linear spin-wave theory. 
We have actually checked that  for $J'/J<-1.25$,  the spectrum $\omega_{\nu,\bm q}$ formally becomes complex around the K$'$ point.

\begin{figure}[bt]
\centering\includegraphics[width=8cm]{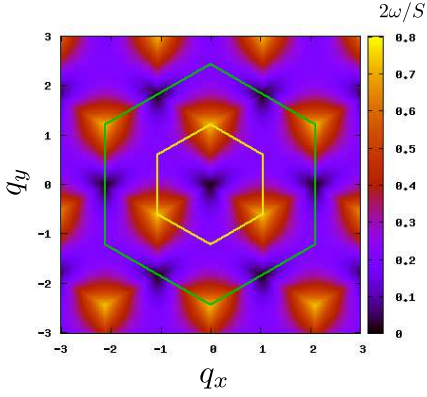}
\caption{(Color online)
The lowest-energy branch of spin-wave dispersion relations for $J=1.0$ and $J'=-1.0$.
The outer(green) hexagon represents the Brillouin zone of the planar Kagome-triangular lattice.
}\label{fig6}
\end{figure}

As in Fig. \ref{fig2}, the transition line between the cuboc and incommensurate orders for the classical case is given by $J=-J'$. 
However, we observe no singular behavior at the classical transition point of $J'/J=-1.0$, and the cuboc phase boundary obtained by the spin-wave theory is extended up to $J'/J=-1.25$.
This is an interesting feature of the non-coplanar spin order, where the geometric phase factor $e^{\pm i \phi}$ may induce a nontrivial quantum effect even at the linear-spin-wave level\cite{relative_angle}.

Here, we would like to comment on a connection to  Domenge's spin-wave result in Ref. \citen{Domenge}, where the  same cuboc spin configuration for a similar Kagome spin system with the nearest-neighbor couplings was investigated. 
A difference between the Kagome-triangular lattice and  Domenge's model in  Ref. \citen{Domenge} is in the connectivity of the next-nearest-neighbor coupling (in the sense of the Kagome lattice);
in the present Kagome-triangular lattice,   half of the nearest neighbor sites are connected with the $J$ coupling.
Then, the hopping paths among  sublattices A, B, C, and D in Fig. \ref{fig3} have one-to-one correspondence to the triangles of the tetrahedron  in the spin configuration space.
Meanwhile, for Domenge's model in Ref. \citen{Domenge},  all of the nearest-neighbor sites are connected with each other.
Then,  we can define  the 120$^\circ$ structure with the opposite vector spin chirality for the surface triangle of the tetrahedron corresponding to  the other half of the nearest-neighbor bonds [e.g., \{2,4,12\} spins in Fig. \ref{fig3}(b)]. 
The cuboc orders for these two models may  be adiabatically connected with each other. 
However, the Kagome-triangular lattice may contain the minimal couplings to realize the cuboc order.
We also note that in Ref. [\citen{Domenge}], the rotating frame based on the triple-$\bm q$ structure is used to reduce the matrix dimension of the spin-wave Hamiltonian, where the role of the geometric phase is not visible in the Hamiltonian level.

\section{Spin-Wave Analysis for the Coupled Spin Tubes}
\label{sec_3d}

We investigate the coupled spin tube system (\ref{chamiltonian}) with full 3D couplings, which can also be regarded as the stacked Kagome-triangular system.
Since there is no frustration along the $c$-axis direction,  the staggered pattern of the cuboc order can be assumed.
Then, the magnetic unit cell contains 24 spins, implying that we deal with 24 kinds of Holstein-Primakoff bosons.
We write a vector array of the bosons for the staggered cuboc order as
\begin{align}
X_{\bm q}^\dagger & =(A^\dagger_{\bm q},B^\dagger_{\bm q},C^\dagger_{\bm q},D^\dagger_{\bm q}, \bar{A}^\dagger_{\bm q}, \bar{B}^\dagger_{\bm q},\bar{C}^\dagger_{\bm q}, \bar{D}^\dagger_{\bm q},\nonumber\\
 &A_{-\bm q}^t, B_{-\bm q}^t, C_{-\bm q}^t, D_{-\bm q}^t, \bar{A}_{-\bm q}^t, \bar{B}_{-\bm q}^t, \bar{C}_{-\bm q}^t, \bar{D}_{-\bm q}^t) ,
\end{align}
where $\bar{A}_{\bm q}, \bar{B}_{\bm q}, \cdots $ denote sets of bosons corresponding to the staggered spins of $A_{\bm q}, B_{\bm q}, \cdots$, respectively.
Then, the spin-wave Hamiltonian can be written as
\begin{align}
{\cal H}^{\rm 3D}_S = \frac{S}{8} \sum_{\bm q} X^\dagger_{\bm q}{\rm H}^{\rm 3D}_{\bm q} X_{\bm q},
\label{3dHl}
\end{align}
where ${\rm H}^{\rm 3D}_{\bm q}$ is a $48\times 48$ matrix.
For the coupled spin tubes, tedious but straightforward calculations yield  
\begin{align}
{\rm H}^{\rm 3D}_{\bm q}=
\begin{pmatrix}
P_{\bm q} & 0 & Q_{\bm q}& Q^c_{\bm q} \\
0 & \bar{P}_{\bm q} & Q^c_{\bm q} & {Q}^\dagger_{\bm q} \\
Q_{\bm q}^\dagger &  Q^c_{\bm q}& \bar{P}_{\bm q} & 0\\
 Q^c_{\bm q} & {Q}_{\bm q} & 0& P_{\bm q} \\
\end{pmatrix}
, \label{elm3dHl}
\end{align}
where ${\bm q}$ is the momentum in the 3D reciprocal lattice vector space.
In this equation, $P_{\bm q}$, $\bar{P}_{\bm q}$, and  $Q_{\bm q}$ are respectively the same as Eqs. (\ref{Pmatrix}), (\ref{Pbarmatrix}), and (\ref{Qmatrix}) with
$K$ replaced by $ K^{\rm 3D}$.
In addition, $Q^c$ represents inter-layer hoppings of the bosons in the $c$-axis direction, which are given by the real diagonal matrix
\begin{align}
Q^c_{\bm q}=
\begin{pmatrix}
\beta_{\bm q} & 0 & 0 & 0 \\
0 & \beta_{\bm q} & 0 & 0 \\
0 & 0 & \beta_{\bm q} & 0\\
0 & 0 & 0 & \beta_{\bm q} \\
\end{pmatrix}
.
\end{align}
Here, $\beta_{\bm q}$ is a $3 \times 3 $ diagonal matrix, which is also given in Appendix A.

The cuboc spin configurations in the adjacent Kagome-triangular layers have the staggered  structure.
This suggests that the matrix elements in Eq. (\ref{elm3dHl})  basically also have the same structure within each layer.
However,  the relative angle when rotating the quantization axis in the cuboc configuration acquires the opposite sign due to the spin inversion, implying that the signs of the geometric phase factors for the two adjacent layers are also alternating.
Here, it should be remarked that the vector chirality on the unit triangle is invariant under the spin inversion, which suggests that the sign of the geometric phases/scalar spin chirality rather than the vector spin chirality is intrinsic to the non-coplanar spin structure of the cuboc order.

\subsection{Spin-wave dispersion relation}

Let us discuss the spin-wave dispersion for the coupled spin tubes.
The numerical Bogoliubov transformation for Eq. (\ref{3dHl}) is straightforwardly constructed as in the 2D case.
In the following, the exchange coupling in the $c$-axis direction is fixed at $J_c=1.0$.
In Fig. \ref{fig7}, we  show spin-wave dispersion relations with $q_c =0$ fixed, where the momentum path is the same as that in Fig. \ref{fig4}.
This is because the dispersion relations in the $c$-direction are basically described by the usual curves for the staggered spin configurations.

\begin{figure}[tb]
\centering\includegraphics[width=6.1cm]{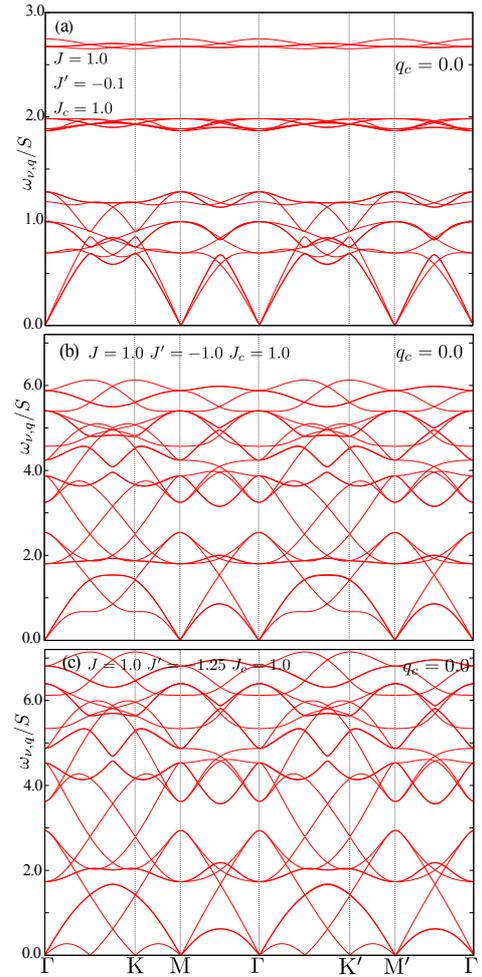}
\caption{(Color online)
Spin-wave dispersion relations of the coupled spin tubes with $J=1.0$ and $J_c=1.0$ in the $q_c=0$ sector:
(a) $J'=-0.1$, (b) $J'=-1.0$, and (c) $J'=-1.25$.
The horizontal axis is along the momentum path defined in Fig. \ref{fig4}.
Note that the scale of the vertical axis in (a) is different from those in (b) and (c).
}\label{fig7}
\end{figure}

In Fig. \ref{fig7}, the spin-wave dispersion relations have soft modes at the $\Gamma$, M, and M$'$ points in consistent with the 2D case.
An important difference from the 2D result is that the K and K$'$ points turn out be equivalent for the 3D case.
As was mentioned above, the staggered configuration of the cuboc order has the opposite sign to the geometric phase, for which the roles of the K and K$'$ points are alternated.
Then, the $q_c=0$ mode for the staggered cuboc order is described by the superposition of the spin waves attributed to the staggered configurations, implying that the K and K$'$ points in Fig. \ref{fig7} become equivalent at the dispersion level.

Another important feature is that in Figs. \ref{fig7}(b) and (c),  the higher energy modes at the K and K$'$ points are lowered, as in the 2D result. 
In particular, we find that these modes touche the zero energy at $J'/J=-1.25$ in Fig. \ref{fig7}(c), which is the same phase boundary as in the 2D case.
 However, the low-energy behavior in the vicinity of the K or K$'$ point is modified to $\omega_{\nu,\bm q}\sim |\bm q -\bm q_{{\rm K/K}'}|$.
In addition, the formal solution of $\omega_{\nu, \bm q}$ for $J'<-1.25$ abruptly becomes complex around the K/K$'$ point, where the staggered cuboc order is unstable.
Thus, it can be concluded that the transition to the incommensurate phase for the 3D case is modified to the first order within the linear spin-wave level.

\subsection{Dynamical spin structure factor}

We discuss the dynamical spin structure factors for the coupled spin tubes, which are relevant to neutron scattering experiments.
In this paper, we consider the dynamical spin structure factor for the net spins in the magnetic unit cell, which is defined as
\begin{equation}
\tilde{S}^{xx}({\bm q},\omega)\equiv \int_{-\infty}^{\infty}\langle \tilde{S}^x_{\bm q}(0) \tilde{S}^x_{-\bm q}(t) \rangle e^{-i\omega t} dt ,
\label{sqw}
\end{equation}
where $ \tilde{S}^x_{\bm q} \equiv \sum_{\nu=1}^{24} S^x_{\nu, \bm q}$ and $\langle \cdots \rangle $ indicates the canonical ensemble average.
The $x$ direction of the spin space is defined in Fig. \ref{fig3}(b).
We also calculate $\tilde{S}^{yy}(\bm q,\omega)$ and  $\tilde{S}^{zz}(\bm q,\omega)$. 
Technical details are briefly summarized in Appendix B.
Below, we present results for a finite temperature, $T/J=0.1$.
However,  note that the results for $T/J=0.1$ are qualitatively the same as those for the ground state. 

\begin{figure}[bt]
\centering\includegraphics[width=7cm]{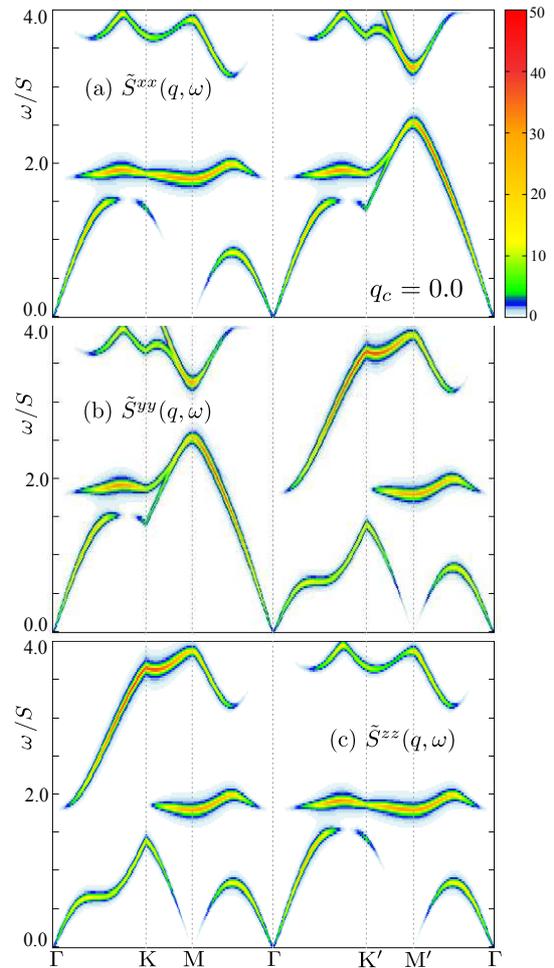}
\caption{(Color online)
Dynamical spin structure factors for the coupled spin tubes of $J=1.0$, $J'=-1.0$, and $J_c=1.0$ at $T/J=0.1$: (a) $S^{xx}(\bm q, \omega)$, (b) $S^{yy}(\bm q,\omega)$, and (c) $S^{zz}(\bm q,\omega)$. 
The relative value of the color-map intensity is meaningful.
}\label{fig8}
\end{figure}

Figure \ref{fig8} shows dynamical spin structure factors in the $q_c=0$ sector for $J=1.0$, $J'=-1.0$ and $J_c=1.0$.
In Fig. \ref{fig8}(a), we can see that $\tilde{S}^{xx}(\bm q,\omega)$ captures  partial branches of the dispersion curves in Fig. \ref{fig7}(b), although the $\Gamma$, M, and M$'$ points are equivalent at the dispersion level.
Also, $\tilde{S}^{yy}(\bm q,\omega)$ in Fig. \ref{fig8}(b) exhibits strong intensity at branches different from $\tilde{S}^{xx}(\bm q,\omega)$.
A similar situation can be seen for $\tilde{S}^{zz}(\bm q, \omega)$ in Fig. \ref{fig8}(c), where modes shifted from those in $\tilde{S}^{xx/yy}(\bm q,\omega)$ are observed.
This is partly because the dispersion curves  with the triple $\bm q$ contain the four modes folded by the translational symmetry breaking  in the $ab$-plane.
In addition, for example, the cuboc spin configurations have anisotropic behaviors in the $x$ and $y$ directions, where the $\pi/2$-rotation with respect to the $z$ axis is not compatible with the cuboc order in Fig. \ref{fig3}(b).
We think that the symmetries in the real space and the spin space may cause such complex selection rules in the dynamical spin structure factors.

In experimental situations,  the directions of the spin configuration space are difficult to control.
In particular, CsCrF$_4$ is synthesized as powder samples. 
 Thus, the angle-averaged dynamical structure factors may be relevant for actual experiments.
We also comment that the strong intensity observed for higher-energy branches may be a characteristic feature for detecting the cuboc order in experiments.

As was seen for the dispersion relations in Fig. \ref{fig7}, the softening at the K or K$'$ points is characteristic to the transition to the incommensurate order. 
In Fig. \ref{fig8}, however, $\tilde{S}^{xx}(\bm q, \omega)$  has no intensity for the lowest-energy branch at the K and K$'$ points, implying that $\tilde{S}^{xx}(\bm q, \omega)$ is not appropriate for detecting the transition point.
Meanwhile,  $\tilde{S}^{yy}(\bm q, \omega)$ or  $\tilde{S}^{zz}(\bm q, \omega)$ have strong intensity at the midpoint of the $\Gamma$ and  K (K$'$) points [i.e. K (K$'$) point of the inner hexagon in Fig.\ref{fig4}].
Thus, we should see  $\tilde{S}^{yy/zz}(\bm q, \omega)$ rather than $\tilde{S}^{xx}(\bm q, \omega)$ to capture the signature of the transition to the incommensurate phase.

\section{Summary and Discussion}

We have investigated low-energy excitations for coupled spin tubes with the cuboc order.
First, we have pointed out the importance of the Kagome-triangular lattice structure in the $ab$-plane.
We then  performed the spin-wave analysis for the 2D Kagome triangular spin system to extract spin-wave excitations, where the soft modes appear at the $\Gamma$, M, and M$'$ points, reflecting the triple-$\bm q$ structure of the cuboc order.
We have also found that the classical phase boundary between the cuboc and incommensurate ordered phases is shifted up to $J'/J=-1.25$, where the geometric phase  characteristic to the cuboc spin structure is essential.
It is an interesting remaining problem to clarify how the spin-wave interaction of higher-$S$ terms affects the present linear-spin-wave results.

We have further examined the spin-wave excitations for the coupled spin tubes or equivalently for the stacked Kagome-triangular system, assuming the staggered cuboc order in the $c$-axis direction. 
As shown in Fig. \ref{fig7}, the spin-wave excitations in the $q_c=0$ sector also have soft modes at the $\Gamma$, M and M$'$ points.
In contrast to the 2D case, the spin-wave dispersion relations show symmetric behavior between the K and K$'$ points because of the contributions from the adjacent staggered layers where the sign of the geometric phase factors is also alternating.
Finally, we have calculated the dynamical spin structure factors $\tilde{S}^{\alpha\alpha}(\bm q, \omega)$ with $\alpha \in \{x,\; y,\; z\}$ for the coupled spin tubes.
We have then found that $\tilde{S}^{\alpha\alpha}(\bm q, \omega)$ have nontrivial $x$-, $y$-,  and $z$-dependence,  reflecting the anisotropy of the cuboc order in the spin configuration space.
In experimental situations, thus, we should carefully take account of the direction dependences of $\tilde{S}^{\alpha\alpha}(\bm q, \omega)$.

Finally, we would like to comment on the relevance to experiments associated with the cuboc order.
As was mentioned, a neutron scattering experiment on CsCrF$_4$ suggests the existence of a nontrivial order below 4 K, but its details have not been identified yet\cite{Hagihala}.
We believe that the present spin-wave results for the cuboc order provide an important reference to identify the order of CsCrF$_4$.
In addition, we should note that CsCrF$_4$ may contain the Dzyaloshinskii-Moriya interaction\cite{CsCrF4_Manaka_2009,DM}.
Thus, how the Dzyaloshinskii-Moriya interaction affects the present spin-wave results is an important problem.
We also think that our spin-wave results could be helpful for analyzing interesting low-energy behaviors observed for planar Kagome-based spin systems such as  NaBa$_2$Mn$_3$F$_{11}$\cite{Ishikawa} and  Cu$_3$Zn(OH)$_6$Cl$_2$\cite{Kapella}.

\begin{acknowledgments}
We would like to thank Y. Akagi and H. Katsura for valuable discussions.
This work was supported by Grants-in-Aid 26400387, 16J02724, and 17H02931 from the Ministry of Education, Culture, Sports, Science and Technology of Japan. 
\end{acknowledgments}

\appendix

\section{Matrix Elements}

We explicitly write down matrices for the spin-wave Hamiltonians in \S \ref{sec_planar} and \S \ref{sec_3d}.
The matrices correspond to three spins in the triangle unit.
Below, $\bm r_1$, $\bm r_2$, and $\bm r_3$ represent the vectors indicating the nearest-neighboring sites in Fig. \ref{fig1}(b),  $\bm r_c$ is the unit vector of the lattice translation in the $c$-axis (tube-leg) direction for the 3D case, and $\bm q$ denotes the momentum.
The phase factor $e^{\pm i \phi}$ originates from the geometric phase defined by the angle between two triangle planes of the tetrahedron.
\begin{align}
K =4
\begin{pmatrix}
J - 2J' & 0 & 0 \\
0 & J - 2J' & 0 \\
0 & 0 & J - 2J'
\end{pmatrix}
\nonumber
\end{align}

\begin{align}
K^{\rm 3D} =4
\begin{pmatrix}
J - 2J' +J_c & 0 & 0 \\
0 & J - 2J'+J_c & 0 \\
0 & 0 & J - 2J' +J_c
\end{pmatrix}
\nonumber
\end{align}
\begin{align}
\alpha_{\bm q}
&=
\begin{pmatrix}
0 & J e^{-i {\bm q \cdot} {\bm r}_3} & Je^{-i {\bm q} \cdot {\bm r}_2} \\
Je^{i {\bm q} \cdot {\bm r}_3} & 0 & Je^{-i {\bm q} \cdot {\bm r}_1} \\
Je^{i {\bm q} \cdot {\bm r}_2} & J^{i {\bm q} \cdot {\bm r}_1} & 0
\end{pmatrix}
\nonumber
\\
\bar{\alpha}_{{\bm q}}
&=
\begin{pmatrix}
0 & Je^{-i {\bm q} \cdot {\bm r}_3} \color{red}{e^{i \phi}} & Je^{-i {\bm q} \cdot {\bm r}_2} \color{red}{e^{i \phi}} \\
Je^{i {\bm q} \cdot {\bm r}_3} \color{red}{e^{i \phi}} & 0 & Je^{-i {\bm q} \cdot {\bm r}_1} \color{red}{e^{i \phi}} \\
Je^{i {\bm q} \cdot {\bm r}_2} \color{red}{e^{i \phi}} & J^{i {\bm q} \cdot {\bm r}_1} \color{red}{e^{i \phi}} & 0
\end{pmatrix}
\nonumber
\end{align}
\begin{align}
\label{matrix}
\beta_{\bm q} =4
\begin{pmatrix}
J_c \cos({\bm q}\cdot{\bm r_c}) & 0 & 0 \\
0 & J_c \cos({\bm q}\cdot{\bm r_c})  & 0 \\
0 & 0 & J_c \cos({\bm q}\cdot{\bm r_c}) 
\end{pmatrix}
\nonumber
\end{align}
\begin{align}
\gamma_{1 \, {\bm q}}
&=
\begin{pmatrix}
0 & 0 & J' e^{i {\bm q} \cdot {\bm r}_1} \color{red}{e^{i \phi}} \\
0 & 0 & J' e^{i {\bm q} \cdot {\bm r}_2} \color{red}{e^{i \phi}} \\
J' e^{-i {\bm q} \cdot {\bm r}_1} \color{blue}{e^{-i \phi}} & J' e^{-i {\bm q} \cdot {\bm r}_2} \color{blue}{e^{-i \phi}} & 0
\end{pmatrix}
\nonumber \\
\gamma_{2 \, {\bm q}}
&=
\begin{pmatrix}
0 & J' e^{-i {\bm q} \cdot {\bm r}_1} \color{red}{e^{i \phi}} & 0 \\
J' e^{i {\bm q} \cdot {\bm r}_1} \color{blue}{e^{-i \phi}} & 0 & J' e^{-i {\bm q} \cdot {\bm r}_3} \color{blue}{e^{-i \phi}} \\
0 & J' e^{i {\bm q} \cdot {\bm r}_3} \color{red}{e^{i \phi}} & 0
\end{pmatrix}
\nonumber \\
\gamma_{3 \, {\bm q}}
&=
\begin{pmatrix}
0 & J' e^{i {\bm q} \cdot {\bm r}_2} \color{blue}{e^{-i \phi}} & J' e^{i {\bm q} \cdot {\bm r}_3} \color{blue}{e^{-i \phi}} \\
J' e^{-i {\bm q} \cdot {\bm r}_2} \color{red}{e^{i \phi}} & 0 & 0 \\
J' e^{-i {\bm q} \cdot {\bm r}_3} \color{red}{e^{i \phi}} & 0 & 0
\end{pmatrix}
\nonumber
\end{align}
\begin{align}
\bar{\gamma}_{1 \, {\bm q}}
&=
\begin{pmatrix}
0 & 0 & J' e^{i {\bm q} \cdot {\bm r}_1} \color{blue}{e^{-i \phi}} \\
0 & 0 & J' e^{i {\bm q} \cdot {\bm r}_2} \color{blue}{e^{-i \phi}} \\
J' e^{-i {\bm q} \cdot {\bm r}_1} \color{red}{e^{i \phi}} & J' e^{-i {\bm q} \cdot {\bm r}_2} \color{red}{e^{i \phi}} & 0
\end{pmatrix}
\nonumber\\
\bar{\gamma}_{2 \, {\bm q}}
&=
\begin{pmatrix}
0 & J' e^{-i {\bm q} \cdot {\bm r}_1} \color{blue}{e^{-i \phi}} & 0 \\
J' e^{i {\bm q} \cdot {\bm r}_1} \color{red}{e^{i \phi}} & 0 & J' e^{-i {\bm q} \cdot {\bm r}_3} \color{red}{e^{i \phi}} \\
0 & J' e^{i {\bm q} \cdot {\bm r}_3} \color{blue}{e^{-i \phi}} & 0
\end{pmatrix}
\nonumber \\
\bar{\gamma}_{3 \, {\bm q}}
&=
\begin{pmatrix}
0 & J' e^{i {\bm q} \cdot {\bm r}_2} \color{red}{e^{i \phi}} & J' e^{i {\bm q} \cdot {\bm r}_3} \color{red}{e^{i \phi}} \\
J' e^{-i {\bm q} \cdot {\bm r}_2} \color{blue}{e^{-i \phi}} & 0 & 0 \\
J' e^{-i {\bm q} \cdot {\bm r}_3} \color{blue}{e^{-i \phi}} & 0 & 0
\end{pmatrix}
\nonumber
\end{align}
\begin{align}
\eta_{ 1\, {\bm q}}
&=
\begin{pmatrix}
0 & 0 & J' e^{i {\bm q} \cdot {\bm r}_1} \\
0 & 0 & J' e^{i {\bm q} \cdot {\bm r}_2} \\
J' e^{-i {\bm q} \cdot {\bm r}_1} & J' e^{-i {\bm q} \cdot {\bm r}_2} & 0
\end{pmatrix}
\nonumber \\
\eta_{2 \, {\bm q}}
&=
\begin{pmatrix}
0 & J' e^{-i {\bm q} \cdot {\bm r}_1} & 0 \\
J' e^{i {\bm q} \cdot {\bm r}_1} & 0 & J' e^{-i {\bm q} \cdot {\bm r}_3} \\
0 & J' e^{i {\bm q} \cdot {\bm r}_3} & 0
\end{pmatrix}
\nonumber \\
\eta_{3 \, {\bm q}}
&=
\begin{pmatrix}
0 & J' e^{i {\bm q} \cdot {\bm r}_2} & J' e^{i {\bm q} \cdot {\bm r}_3} \\
J' e^{-i {\bm q} \cdot {\bm r}_2} & 0 & 0 \\
J' e^{-i {\bm q} \cdot {\bm r}_3} & 0 & 0
\end{pmatrix}
\nonumber
\end{align}
Note that $\bar{\alpha}_{\bm q}$ is {\it not} Hermitian.

\section{ Calculation of $S(\bm q , \omega)$}

We briefly summarize computational details of the dynamical structure factor for the net spins in the magnetic unit cell, which is defined by Eq. (\ref{sqw}).
Using the Holstein-Primakoff bosons, we have 
\begin{align}
\tilde{S}^x_{\bm q} =\sum_{\nu}\xi_\nu a_{\nu, \bm q} +\xi_{\nu}^* a^\dagger_{\nu, -\bm q}+  \zeta_\nu (S-a^\dagger_{\nu, \bm q}a_{\nu,\bm q}) 
\end{align}
within  ${\cal O}(S^{0})$.
Here, $\xi$ contains  coefficients of ${\cal O}(S^{1/2})$ due to the Holstein-Primakoff transformation as well as geometric coefficients attributed to the orientation of spins in the magnetic unit cell.
Meanwhile, $\zeta$ is determined only by the orientation angle of spins.
In the following, moreover, we assume that the average spin of the classical order in the magnetic unit cell is zero, implying 
\begin{align}
\sum_\nu \zeta_\nu =0,
\end{align}
which is actually the case for the cuboc order.
Then, it is sufficient to consider
\begin{align}
\tilde{S}^x_{\bm q} \simeq \sum_{\nu}\xi_\nu a_{\nu, \bm q} +\xi^*_{\nu} a^\dagger_{\nu, -\bm q}
\end{align}
for the calculation  of $S(\bm q , \omega)$ up to ${\cal O}(S)$.
Introducing a vector array of the coefficients,
\begin{align}
{\rm v}^t= (\xi_1,\cdots, \xi_1^*,\cdots),
\end{align}
we may write
\begin{align}
\tilde{S}^x_{\bm q}(0)\tilde{S}^x_{-\bm q}(t) = \left( X_{\bm q}^t(0) {\rm v} \right)\left({\rm v}^t  X_{-\bm q}(t)\right) .
\label{sqs-q}
\end{align}
Using the relation
\begin{align}
X^t_{\bm q}(0) {\rm v}=  X^\dagger_{- \bm q}(0) {\rm v}^* ,
\end{align}
we can write Eq. (\ref{sqs-q}) as the standard form, 
\begin{align}
\tilde{S}^x_{\bm q}(0)\tilde{S}^x_{-\bm q}(t) = X^\dagger_{-\bm q}(0) {\rm L}^{xx} X_{-\bm q}(t), 
\end{align}
where ${\rm L}^{xx} \equiv {\rm v}^*{\rm v}^t$.
Using the Bogoliubov transformation $X_{\bm q} = {\rm T}_{\bm q} X'_{\bm q}$,  we then have
\begin{align}
\langle \tilde{S}^x_{\bm q}(0)\tilde{S}^x_{-\bm q}(t)\rangle
 = \langle X'^\dagger_{-\bm q}(0) {\rm M}^{xx}_{-\bm q}   X'_{-\bm q}(t) \rangle,
\label{average-m}
\end{align}
where ${\rm M}^{xx}_{-\bm q} \equiv {\rm T}_{-\bm q}^\dagger {\rm L}^{xx}{\rm T}_{-\bm q}$ is Hermitian.
Here, $X'_{-\bm q}$ defines Bogoliubov particles diagonalizing the Hamiltonian such as $\alpha_{\nu, -\bm q}$ and  $\alpha^\dagger_{\nu, \bm q}$.
In Eq. (\ref{average-m}), the diagonal elements of ${\rm M}^{xx}_{-\bm q}$ contribute to the average.
 The Heisenberg equation of motion yields $\alpha_{\nu, \bm q}(t)=\alpha_{\nu, \bm q} e^{-i\omega_{\nu, \bm q}t}$, etc. 
Then, we can perform the $t$ integral in Eq. (\ref{sqw})  and arrive at
\begin{align}
& S^{xx}({\bm q},\omega_{\bm q}) = \pi \sum_\nu [{\rm M}^{xx}_{-\bm q}]_{\nu,\nu} f_B(\beta \omega_{\nu,-\bm q}) \delta(\omega+\omega_{\nu, - \bm q}) \nonumber \\
& +\pi \sum_\nu [{\rm M}^{xx}_{-\bm q}]_{N+\nu,N+\nu} (f_B(\beta \omega_{\nu,-\bm q})+1) \delta(\omega-\omega_{\nu,- \bm q})),
\label{sqw_res}
\end{align}
where $f_B$ is the Bose distribution function.
This expression is useful for numerical computations.
Note that in Fig. \ref{fig8}, we approximate the $\delta$-function in Eq. (\ref{sqw_res}) with the Lorentzian of the broadening factor $\epsilon = 0.02$.


\begin{thebibliography}{99}
\bibitem{relative_angle} If we omit the phase factor $e^{\pm i \phi}$ in Eq. (\ref{2dHl}), the phase boundary  corresponds to the classical one, $J'/J=-1.0$.
\end{thebibliography}
\end{document}